\documentclass[eqsecnum,preprint,prd,aps,nofootinbib]{revtex4}
\usepackage{amsmath}
\usepackage{graphics}
\newcommand{\be}{\begin{equation}}
\newcommand{\ee}{\end{equation}}
\newcommand{\ba}{\begin{eqnarray}}
\newcommand{\ea}{\end{eqnarray}}
\newcommand{\bc}{\begin{center}}
\newcommand{\ec}{\end{center}}

\begin{document}
\begin{center}
\bibliographystyle{article}

{\Large \textsc{Noether symmetry approach in pure gravity with variable
$G$ and $\Lambda$}}

\end{center}
\vspace{0.4cm}


\date{\today}

\author{Alfio Bonanno,$^{1,2}$ \thanks{%
Electronic address: abo@ct.astro.it} Giampiero Esposito$^{3,4}$ \thanks{%
Electronic address: giampiero.esposito@na.infn.it} Claudio Rubano$^{4,3}$
\thanks{%
Electronic address: claudio.rubano@na.infn.it} and Paolo Scudellaro$^{4,3}$
\thanks{%
Electronic address: scud@na.infn.it}}
\affiliation{${\ }^{1}$Osservatorio Astrofisico, Via S. Sofia 78, 95123 Catania, Italy\\
${\ }^{2}$Istituto Nazionale di Fisica Nucleare, Sezione di Catania,\\
Corso Italia 57, 95129 Catania, Italy\\
${\ }^{3}$Istituto Nazionale di Fisica Nucleare, Sezione di Napoli,\\
Complesso Universitario di Monte S. Angelo, Via Cintia, Edificio N', 80126
Napoli, Italy\\
${\ }^{4}$Dipartimento di Scienze Fisiche, Complesso Universitario di Monte
S. Angelo,\\
Via Cintia, Edificio N', 80126 Napoli, Italy}

\begin{abstract}
We find exact cosmological solutions when the Newton parameter and
the cosmological term are dynamically evolving in a
renormalization-group improved Hamiltonian approach. In our
derivation we use the Noether symmetry approach, leading to an
interesting variable transformation which yields exact and
general integration of the cosmological equations. The functional dependence 
of $\Lambda$ on $G$ is determined by the
method itself, therefore generalizing previous results on symmetry
principles in cosmology. We find new functional relations between 
$\Lambda$ and $G$, jointly with power-law inflation for pure gravity.
\end{abstract}
\maketitle
\bigskip
\vspace{2cm}

\section{Introduction}

Among the most important questions in cosmology, the task of 
understanding the origin and the nature of the Big Bang still remains closely
tied to unavoidable problems of physical description and
interpretation. This is fully connected to the unexploited
marriage between classical general relativity and quantum field
theory. Many have been the proposals for solving the several
issues arising from such an effort. An original approach suggests
the possibility that the cosmological dynamics is generated by
strong ``renormalization group (hereafter RG) induced'' quantum effects
which would drive the (dimensionless) cosmological ``constant''
$\lambda(k)$ and Newton ``constant'' $g(k)$ from an ultraviolet
attractive fixed point.

Within the framework of the effective average action
\cite{Berg02,Wett01,Reut94} a suitable fixed point is in fact
known to exist in the Einstein--Hilbert truncation of theory space
\cite{Reut98,Laus02a,Soum99} and in the higher-derivative
generalization \cite{Laus02b}. There are indications
\cite{Laus02a,Reut02b,Laus02b,Bona05,Nied03,Nied02,Forg02} that a
non-Gaussian ultraviolet fixed point should indeed exist in the
exact theory, implying its non-perturbative renormalizability.
Within this RG-improved framework, gravitational phenomena at a
typical distance scale $\ell\equiv k^{-1}$ can be described in
terms of a scale-dependent effective action
$\Gamma_k[g_{\mu\nu}]$, thought of as a Wilsonian coarse-grained
free-energy functional which has been identified \cite{Reut98}
with the effective average action for Euclidean quantum gravity;
an exact functional RG equation for the $k$-dependence of
$\Gamma_k$ has also been derived. Non-perturbative solutions were
then obtained in such a context, referred to as \emph{quantum
Einstein gravity}. The RG equations offer an explicit answer for
the $k$-dependence of the running Newton term $G(k)$ and the
running cosmological term $\Lambda(k)$, which is very important
for an understanding of the Planck era immediately after the big
bang and the structure of the black hole singularity
\cite{Bona02b,Bona99,Bona00}.
The RG-improved Einstein equations for a homogeneous and
isotropic universe are obtained, for example, identifying $k$ with
the inverse of cosmological time, $k \propto 1/t$
\cite{Bona02b,Bona02a} and by promoting $G$ and $\Lambda$ to 
dynamically evolving quantities in the Einstein equations. 

This approach has been further investigated by Reuter and Weyer
in \cite{Reut04}, where an RG-improvement at the level of the action
has been proposed. In particular, it has been shown that a new 
effective interaction of the type $\nabla_{\nu}G \; \nabla_{\mu}G$
should occur in the ``bare'' Lagrangian. It is therefore interesting to
understand the Hamiltonian structure (if any) of the theory when a
kinetic term involving the integral of the inner product of ${\rm grad}G$ 
with itself, divided by $G^{3}$ (see section 2), is present in the
action. This question was addressed in \cite{Bona04}, where it was shown
that a consistent Hamiltonian structure can be obtained by considering
$G$ and $\Lambda$ as dynamical variables instead of regarding them as
external fields. On the other hand, it must always be checked that the
solution is consistent with a realistic RG trajectory. Interestingly,
there exists a class of power-law solutions for the RG-improved
Arnowitt--Deser--Misner (hereafter ADM)
Lagrangian which exactly reproduce the scaling law near the fixed
point for which the product $\Lambda G$ is constant \cite{Tsam93}.

The aim of this paper is to further extend the investigations of 
\cite{Bona04} by using the {\it Noether Symmetry Approach}, a simple
and powerful tool to obtain general solutions of the cosmological 
equations \cite{deritis90, cap96}, and to check their 
consistency with the RG flow
predicted by the $\beta$-functions for the Einstein--Hilbert Lagrangian.
In particular, we shall thus show that it is possible to generate a class
of cosmological solutions consistent with a scaling law of the type
$$
\Lambda G^{2}={\rm constant},
$$
which is realized in the cross-over region close to that Gaussian
fixed point, according to the phase structure described in 
\cite{Reut02b, Reut04, Rreut04}. The resulting cosmology can then be
useful in describing the Early Universe during the Planck Era, before
the classical evolution has taken over \cite{Reut02b}.    

The structure of the paper is as follows:
in section 2 we describe the Lagrangian formulation and find the
Noether symmetry for the Lagrangian adopted to derive the
RG-improved Einstein cosmological equations when the
energy-momentum tensor is vanishing. Section 3 studies how this gives
rise to exact and general solutions, section 4 describes the relation
between Noether symmetries and RG evolution, while concluding
remarks and a critical assessment are presented in sections 5 and 6.

\section{Lagrangian formulation and Noether symmetry}

In \cite{Bona04} the ADM formalism has been applied to models of
gravity with variable $G$ and $\Lambda$. The main foundational steps
therein can be summarized as follows. 
\vskip 0.3cm
\noindent
(i) On renormalization-group improving the gravitational Lagrangian
in the ADM approach, one might think that $G$ and $\Lambda$ have the 
status of given external field, whose evolution is in principle ruled
by the RG flow equation. However, it is instead possible to generalize
the standard ADM Lagrangian and regard $G$ as a dynamical field obeying
an Euler--Lagrange equation, the underlying idea being that all fields
occurring in the Lagrangian $L$ should be ruled by $L$ in the first
place. This makes it possible to fully exploit the potentialities of
the action principle (when $\Lambda$ and $G$ are viewed as external 
fields one is instead halfway through towards such potentialities).
\vskip 0.3cm
\noindent
(ii) For this purpose, one adds to an action of the Einstein--Hilbert
type (but with $G$ variable so that it is brought within the integrand)
two compensating terms such that the action reduces to the 
York--Gibbons--Hawking form for fixed $G$ and $\Lambda$, and takes the
same functional form as the ADM action for general relativity (despite
having variable $G$ and $\Lambda=\Lambda(G)$), 
i.e. ruled by the Lagrangian
$$
L={1\over 16 \pi}\int {N\sqrt{h}\over G}\Bigr(K_{ij}K^{ij}-K^{2}
+{ }^{(3)}R-2\Lambda(G) \Bigr)d^{3}x.
$$
This should be supplemented by an interaction term of a kinetic type,
i.e.
$$
L_{\rm int}=-{\mu \over 16 \pi}\int {g^{\rho \sigma}G_{;\rho}
G_{;\sigma}\over G^{3}}\sqrt{-g}d^{3}x.
$$
Non-vanishing values of $\mu$ ensure that the resulting Euler--Lagrange
or Hamilton equations for $G$ itself are well-defined and admit 
non-trivial solutions. 
\vskip 0.3cm
\noindent
(iii) The functional relation between $\Lambda$ and $G$: 
$\Lambda=\Lambda(G)$ is not an assumption but it follows directly from
the RG equation. Indeed, in general terms, the RG equation reads
\cite{Reut02b}
$$
{\partial G \over \partial k}=\beta_{G}(G(k),\Lambda(k),k),
$$
$$
{\partial \Lambda \over \partial k}
=\beta_{\Lambda}(G(k),\Lambda(k),k),
$$
where $\beta_{G}$ and $\beta_{\Lambda}$ are the $\beta$-functions. Once
the solution $G=G(k)$ and $\Lambda=\Lambda(k)$ is obtained, one can always
eliminate the $k$ dependence in order to obtain the trajectory 
$\Lambda=\Lambda(G)$. In other words, we are saying that we consider
$\Lambda=\Lambda(G)$ along a given RG trajectory, i.e. a solution of the
RG equation. Because of the tremendous technical difficulties we cannot
actually solve the RG equation except near a fixed point, therefore we
assume that a given trajectory exists, and we investigate what type of
dynamics is consistent with a given $\Lambda=\Lambda(G)$ relation. Moreover,
on a quite independent ground, i.e. at Hamiltonian level, we
can point out that, if 
$\Lambda$ were independent of $G$, we would find from the ADM action
the primary constraint of vanishing momentum $\pi_{\Lambda}$ conjugate
to $\Lambda$. The preservation of this primary constraint would lead
to an unacceptable secondary constraint, i.e. a vanishing lapse function    
\cite{Bona04}.

In a homogeneous and isotropic universe, this
approach is shown to lead to power-law behaviours of the cosmic scale
factor $a = a(t)$ for both pure gravity and a massless $\phi^4$
theory, well in agreement with what occurs in fixed-point
cosmology, once one adopts the constraint $\Lambda G = {\rm const}$. 
Here, we want to investigate again the pure-gravity case in
homogeneous and isotropic cosmology (with a signature $-,+,+,+$
for the metric). For this purpose, let us start from
the Lagrangian motivated by the previous considerations \cite{Bona04}
\be
\label{1} L = \frac{a^3}{16 \pi G} \left(
-6\frac{\dot{a}^2}{a^2}+\frac{6{\cal K}}{a^2}-2\Lambda \right) +
\frac{\mu}{16 \pi}\frac{a^3 \dot{G}^2}{G^3}\,,
\ee
where $G=G(t)$,
$\Lambda=\Lambda(G(t))$, ${\cal K} = -1,0,1$ (for open, spatially
flat and closed universes, respectively), and $\mu \neq 0$ is an
interaction parameter on which we do not have any observational
constraint, since it is non-vanishing only for significant
modifications of general relativity, which can indeed occur in the
very early universe; dots indicate time derivatives. $L$ can be
recast in the form \be \label{2} L = \frac{1}{8 \pi G} \left(
-3a\dot{a}^2 + 3{\cal K}a - a^3 \Lambda + \frac{1}{2}\mu a^3
\frac{\dot{G}^2}{G^2} \right)\,, \ee and the resulting
second-order Euler--Lagrange equations for $a$ and $G$ are
\cite{Bona04}
\be \label{3}
\frac{\ddot{a}}{a} +
\frac{\dot{a}^2}{2a^2} + \frac{{\cal K}}{2a^2} - \frac{\Lambda}{2}
- \frac{\dot{a}\dot{G}}{a G} + \frac{\mu \dot{G}^2}{4G^2} =
0\,,\ee \be \label{4} \mu \ddot{G} - \frac{3}{2}\mu
\frac{{\dot{G}}^2}{G} + 3\mu \frac{\dot{a}}{a}\dot{G} +
\frac{G}{2} \left( -6\frac{\dot{a}^2}{a^2} + \frac{6{\cal
K}}{a^2}-2\Lambda + 2G\frac{d\Lambda}{d G} \right) = 0\,.
\ee
By virtue of the Hamiltonian constraint, we have also to
consider the equation \cite{Bona04}
\be \label{5}
\frac{\dot{a}^2}{a^2} + \frac{{\cal K}}{a^2} - \frac{\Lambda}{3} -
\frac{\mu}{6}\frac{\dot{G}^2}{G^2} = 0\,,
\ee
which can indeed be
seen as equivalent to the following constraint on the
\emph{energy} function associated to $L$: 
\be 
\label{5bis} 
E_L \equiv \frac{\partial L}{\partial \dot{a}}\dot{a} + \frac{\partial
L}{\partial \dot{G}}\dot{G} - L = 0\,. 
\ee
Let us point out that this set of equations contains some undetermined
parameters $(\mu,{\cal K})$ and, mostly important, the arbitrary
function $\Lambda(G)$, which plays here a role similar to the potential
$V(\phi)$ in the case of a scalar field. This suggests the possibility
of using the so-called ``Noether symmetry approach'' 
\cite{deritis90, cap96} mentioned in the introduction 
in order to find a possible guess for this
function. On relying upon Refs. \cite{deritis90, cap96} for a complete
discussion of this approach we limit ourselves to say that, in many
circumstances (and here also as we shall see in a moment) this procedure
not only allows a reduction of the dynamical system but leads to 
{\it general exact solutions}, which seems to us a major advantage. On
the other hand, the physical motivation for this choice is just founded
on the search for simplicity and ``beauty'' of the theory. 
Let us first show, therefore,
that with a suitable choice of the function $\Lambda = \Lambda(G)$
a Noether symmetry exists for $L = L(a,G)$, i.e., for the
Lagrangian viewed as a function of $a$ and $G$, considered as
coordinates of the associated configuration space
\cite{deritis90,cap96}.

On considering the vector field
\be \label{6}
X \equiv \alpha
(a,G)\frac{\partial{}}{\partial a} +
\dot{\alpha}\frac{\partial{}}{\partial \dot{a}} + \beta
(a,G)\frac{\partial{}}{\partial G} +
\dot{\beta}\frac{\partial{}}{\partial \dot{G}}\,,
\ee
with $\alpha
= \alpha (a,G)$ and $\beta = \beta (a,G)$
generic $C^{1}$ functions, and
$\dot{\alpha} \equiv d\alpha/dt = (\partial \alpha/\partial
a)\dot{a} + (\partial \alpha/\partial G)\dot{G}$, $\dot{\beta}
\equiv d\beta/dt = (\partial \beta/\partial a)\dot{a} + (\partial
\beta/\partial G)\dot{G}$, we impose the condition 
\be \label{7}
{\cal L}_X L = 0\,,
\ee
${\cal L}_X L$ being the Lie derivative of
$L$ along $X$. This in fact corresponds to a set of equations for
$\alpha = \alpha (a,G)$, $\beta = \beta (a,G)$ and $\Lambda =
\Lambda (G)$
\ba G\alpha +
2aG\frac{\partial \alpha}{\partial a} - a\beta & = & 0\,, \\
3G\alpha - 3a\beta + 2aG\frac{\partial \beta}{\partial G} & = &
0\,,
\\ 6G^2\frac{\partial \alpha}{\partial G} -
\mu a^2\frac{\partial \beta}{\partial a} & = & 0\,, \\
\frac{3{\cal K}}{a^2} \left( \alpha - \frac{a\beta}{G} \right) -
3\Lambda \alpha - a \frac{d\Lambda}{d G} \beta + \frac{a\Lambda
\beta}{G} & = & 0\,. \ea 
Such a system is overdetermined, and we
can in fact use only the first three equations to get the
expressions of $\alpha = \alpha (a,G)$, $\beta = \beta (a,G)$ and,
afterwards, use the fourth one as a consistency equation to
constrain the function $\Lambda = \Lambda (G)$ and, possibly, the
values of ${\cal K}$. As a matter of fact, a solution of Eqs.
(2.9), (2.10) and (2.11) can be found by making the simple
\emph{ansatz} $\alpha \equiv A_1(a)A_2(G)$, which plugged into Eq.
(2.9) constrains the form of $\beta$ to be $\beta \equiv
B_1(a)B_2(G)$, too. On using also Eqs. (2.10) and (2.11), and getting
rid of unnecessary integration constants, we find 
\ba
\alpha (a,G) & \equiv & A_1(a)A_2(G) = a^{\frac{J}{3-2J}}G^{J-1}\,, \\
\beta (a,G) & \equiv & B_1(a)B_2(G) =
\frac{3}{3-2J}a^{\frac{3(J-1)}{3-2J}}G^J\,, 
\ea 
where $J$ is completely fixed in terms of the parameter $\mu$ that
appears in the action, since 
\be \label{13bis} \mu =
\frac{2}{3}(3-2J)^2\,. 
\ee 
From Eqs. (2.14) and (2.15), we thus
get $J \neq 1, 3/2$, so that $\mu \neq 0, 2/3$. As a matter of
fact, from now on we will use the $J$ parameter instead of the
$\mu$ parameter introduced in the Lagrangian $L$, since it makes it
possible to cast the resulting formulae in a more convenient form.

As a consequence, the consistency equation (2.12) becomes
\be
\label{14} \frac{2J{\cal K}}{(2J-3)a^2} = \frac{1}{3-2J}\left[
2(1-J)\Lambda + G\frac{d\Lambda}{d G} \right] 
= {\rm constant} \,\,,
\ee
since the left--hand side depends only on $a$ while the
right--hand side depends only on $G$. This is possible, therefore, if
both
\be \label{15}
J{\cal K} = 0
\ee
and
\be \label{16}
2(1-J)\Lambda + G\frac{d\Lambda}{d G} = 0
\ee
hold, which leads to
the necessity of splitting the next considerations into two
separate branches. As a matter of fact, always suitably neglecting
unnecessary integration constants (except $\Lambda_0$), we find
\be \label{17}
X_0 \equiv X|_{J=0} =
\frac{1}{G}\frac{\partial{}}{\partial a} -
\frac{\dot{G}}{G^2}\frac{\partial{}}{\partial \dot{a}} +
\frac{1}{a}\frac{\partial{}}{\partial G} -
\frac{\dot{a}}{a^2}\frac{\partial{}}{\partial \dot{G}}
\ee
for $J=0$ ($\Rightarrow$ $\mu = 6$), any ${\cal K}$, and 
$$\Lambda =
\Lambda_0 G^{-2},
$$ 
while
\ba
X_J & \equiv & X|_{\forall
J\neq 0} = a^{\frac{J}{3-2J}}G^{J-1}\frac{\partial{}}{\partial a}
+ aG^{J-2}\left[ \frac{J}{3-2J}G\dot{a} +
(J-1)a^{\frac{3(J-1)}{3-2J}}\dot{G} \right]
\frac{\partial{}}{\partial \dot{a}} \nonumber \\  &  & +
\frac{3}{3-2J}a^{\frac{3(J-1)}{3-2J}}G^J\frac{\partial{}}{\partial
G} + \frac{3}{3-2J}a^{\frac{5J-6}{3-2J}}G^{J-1}\left[
\frac{3(J-1)}{3-2J}G\dot{a} + Ja\dot{G} \right]
\frac{\partial{}}{\partial \dot{G}}
\ea
for $J\neq 0, 1, 3/2$
($\Rightarrow$ any $\mu \neq 0, 2/3$), ${\cal K}=0$, and $\Lambda
= \Lambda_0 G^{2(J-1)}$.
Let us stress that $X_J \rightarrow X_0$
for $J \rightarrow 0$, but the situation with $J=0$ cannot indeed
be treated as a mere subcase except when ${\cal K}=0$ ($X$
without any subscript stands for $X_{J}$ for any $J$).
As we said before, this result is a direct consequence of our approach.
Other choices are possible. However, in the case of the scalar field in
standard cosmology, we have found that the functional form of 
$V(\phi)$ is essentially the {\it only one} which allows exact integration
in a simple way \cite{deri90, deri95}. 
It is unknown at present whether the same holds for the
models considered in our paper, and this is a relevant topic for
further research.

\section{Solutions from new coordinates and Lagrangian}

Let us now look for a change of coordinates, such that one of them
is cyclic for the Lagrangian $L$, and the transformed Lagrangian produces
equations that, now, can be easily dealt with. Since
we have proved the existence of a Noether symmetry for $L$,
in fact, we indeed expect that \emph{there is} such a
transformation $\{a,G\} \rightarrow\{u,v\}$, in which we can
assume that $u$ is the new cyclic coordinate, for instance, and
try to deduce it, therefore, by solving the following system of
equations:
\ba
i_{X} d u & = &
a^{\frac{J}{3-2J}}G^{J-1}\frac{\partial u}{\partial a} +
\frac{3}{3-2J}a^{\frac{3(J-1)}{3-2J}}G^J\frac{\partial u}{\partial
G} = 1\,, \\ i_X d v & = & a^{\frac{J}{3-2J}}G^{J-1}\frac{\partial
v}{\partial a} +
\frac{3}{3-2J}a^{\frac{3(J-1)}{3-2J}}G^J\frac{\partial v}{\partial
G} = 0\,,
\ea
$i_X d u$ and $i_X d v$ being the contractions
between the vector field $X$ and the differential forms $d u$ and
$d v$, respectively \cite{deritis90,cap96}. By virtue of the
remarks made at the end of the previous section, we can treat
together all the cases given by different values of $J$.

In order to solve this system, we can again use an \emph{ansatz}
on the forms of $u = u(a,G)$ and $v = v(a,G)$, also noting that, for
our purpose, it is anyway enough to find just \emph{one} solution. Thus,
we can get rid of unnecessary constants, for example, and simply
look for a particular solution. First of all, let us write Eq.
(3.1) as
\be \label{21}
a\frac{\partial u}{\partial a}  +
\frac{3}{3-2J}G\frac{\partial u}{\partial G} =
a^{\frac{3(1-J)}{3-2J}}G^{1-J},
\ee
and let us define
\be \label{22}
u \equiv n\,a^{\frac{3(1-J)}{3-2J}}G^{1-J} + u_0\,,
\ee
$n$ and $u_0$ being two generic constants. This soon solves the
equation (3.3) above, fixing
\be \label{3.5} 
n \equiv n(J) \equiv
\frac{3-2J}{6(1-J)}\,,
\ee
which must be non-vanishing, being $J \neq
3/2$. Using $v \equiv v_1(a) + v_2(G)$, on the other hand, also
solves Eq. (3.2). When $J\neq 0$ (and ${\cal K}=0$), apart from
some integration constants and $u_0$, we have (also setting $m
\equiv m(J) \equiv 1-J$, with $m \neq -1/2, 0$, so that $m \equiv
1/[2(3n - 1)]$) 
\ba 
u = u(a,G) & = &
\frac{3-2J}{6(1-J)}a^{\frac{3(1-J)}{3-2J}}G^{1-J} = n\,
a^{\frac{1}{2n}} G^m\,,
\\ v = v(a,G) & = & \log{\left( a G^{\frac{2J-3}{3}} \right)}
= \log{\left( a G^{-2n m} \right)}\,, 
\ea 
from which we get 
\ba 
u=u(a,G) & = & \frac{1}{2}a G\,, \\ v = v(a,G) & = & \log{\left( a
G^{-1} \right)} 
\ea 
for $J=0$ ($\Rightarrow n = 1/2$) and any ${\cal K}$.

In general, this involves
\ba
a = a(u,v) & = & \left[
\frac{6(1-J)}{3-2J} \right]^{\frac{3-2J}{6(1-J)}} \exp{\left(
\frac{v}{2} \right)} u^{\frac{3-2J}{6(1-J)}} =
n^{-n} \exp{\left( \frac{v}{2} \right)} u^n\,, \\
G = G(u,v) & = & \left[ \frac{6(1-J)}{3-2J}
\right]^{\frac{1}{2(1-J)}} \exp{\left( \frac{3}{2(2J-3)}v \right)}
u^{\frac{1}{2(1-J)}} = {\left[ \frac{1}{n} \exp{\left( -
\frac{v}{2n} \right)} u \right],}^{1/(2m)}
\ea
so that
\ba
a= a(u,v) & = & \sqrt{2u}\exp{\left(
\frac{v}{2} \right)}\,, \\
G = G(u,v) & = & \sqrt{2u}\exp{\left( -\frac{v}{2} \right)},
\ea
when $J=0$ (in such expressions, one has $\mu = 2(3-2J)^2/3 =
6n^2/(1-3n)^2$).

We are now able to deduce the new expressions assumed by the
Lagrangian $L$ when we substitute such functions $a = a(u,v)$ and
$G = G(u,v)$ into it. This in fact produces two different $L'$
in the two separate cases we are describing, that is
\be \label{31}
L'_0 = - 6 \exp{(2v)}\dot{u}\dot{v} + 3{\cal
K}\exp{(v)} - \Lambda_0 \exp{(3v)} 
\ee 
for $J=0$ and any $\cal K$, and 
\be \label{32} 
L'_J = -6\exp{\left( \frac{3(J-2)}{2J-3}v
\right)}\dot{u}\dot{v} - \Lambda_0 \exp{(3v)} = -6\exp{\left(
\frac{m+1}{2n m}v \right)}\dot{u}\dot{v} - \Lambda_0 \exp{(3v)}
\ee
for $J\neq 0$ (and ${\cal K}=0$). As expected, in both cases
$u$ is cyclic for $L'$. This means that there exists a constant of
motion $\Sigma \equiv -\partial L'/\partial \dot{u}$ associated to
$L'$, which makes it possible to solve the equations we can derive from 
this Lagrangian. For each case, in fact, we respectively have
\be \label{33}
\Sigma_0 = 6 \exp{(2v)}\dot{v} \ee and \be \label{34}
\Sigma_J = 6 \exp{\left( \frac{3(J-2)}{2J-3}v \right)} \dot{v} = 6
\exp{\left( \frac{m+1}{2n m}v \right)} \dot{v}\,,
\ee
where $\Sigma_J \rightarrow \Sigma_0$ for $J \rightarrow 0$. As to the
possible values of the constant of motion $\Sigma$, it must
also be non-vanishing, since $\Sigma = 0$ soon yields $\dot{v} = 0$,
which gives no dynamics, as can be seen from the expressions of
$L'$ in Eqs. (\ref{31}) and (\ref{32}). Moreover, it is important to
stress that such new Lagrangians are non-singular, since the
Hessian related to them turns out to be non-vanishing.

Last, to the new Lagrangians $L'$ we can also associate the
two \emph{energy} functions \cite{deritis90,cap96} 
\be \label{35}
E'_0 = - 6 \exp{(2v)}\dot{u}\dot{v} - 3{\cal K}\exp{(v)} +
\Lambda_0 \exp{(3v)} \ee and \be \label{36} E'_J = - 6\exp{\left(
\frac{3(J-2)}{2J-3}v \right)}\dot{u}\dot{v} + \Lambda_0 \exp{(3v)}
= - 6\exp{\left( \frac{m+1}{2n m}v \right)}\dot{u}\dot{v} +
\Lambda_0 \exp{(3v)}\,, 
\ee 
respectively, so that again $E'_J
\rightarrow E'_0$ when $J \rightarrow 0$ and ${\cal K}=0$.

\subsection{Case with $J=0$ and generic ${\cal K}$}

Equation (\ref{33}) can be seen as a first-order
differential equation for $v$ in terms of time $t$ ($\Sigma_0$
becoming a parameter in it), so that its general solution is 
\be \label{37} 
v = v(t) = \frac{1}{2} \log{\left(\frac{\Sigma_0}{3}t +
2C_1 \right)}\,, 
\ee 
where $C_1$ is an arbitrary integration
constant. Equation (\ref{35}) becomes therefore a first-order
differential equation for $u$ in terms of $t$ (and $\Sigma_0$),
and we find \be \label{38} u = u(t) =
\frac{2}{15\sqrt{3}{\Sigma_0}^2}(\Sigma_0 t + 6C_1)^{3/2}(\Sigma_0
\Lambda_0 t - 15{\cal K} + 6C_1 \Lambda_0) + C_2\,, \ee $C_2$
being another arbitrary integration constant. From now on,
we set $C_2=0$. This indeed makes us lose the generality of
our solution, which is on the contrary guaranteed from the
existence of the three integration constants $C_1$, $C_2$, and
$\Sigma_0$, plus the condition $E_L =0$. We set hereafter $C_{2}=0$
to simplify the resulting formulae.

Let us rescale time by defining $\tau \equiv \Sigma_0 t + 6C_1$ (which
tells us that $\Sigma_0$ must be positive), so that 
\be \label{39}
u(\tau) = \frac{2{\tau}^{3/2}(\Lambda_0 \tau - 15{\cal
K})}{15\sqrt{3}{\Sigma_0}^2} \,,\,\,\,\,\, v(\tau) = \frac{1}{2}
\log{\left(\frac{\tau}{3} \right)}\,, 
\ee 
which can be substituted
into Eqs. (3.8) and (3.9), eventually obtaining
\ba
a = a(\tau) & =
& \frac{2\tau}{3\sqrt{5}\Sigma_0}\sqrt{\Lambda_0 \tau - 15{\cal
K}}\,,
\\ G = G(\tau) & = &
\frac{2\sqrt{\tau}}{\sqrt{15}\Sigma_0}\sqrt{\Lambda_0\tau -
15{\cal K}}\,,
\ea
and
\be \label{3.25}
\Lambda = \Lambda(G(\tau))
= \Lambda_0 G^{-2}(\tau) =
\frac{15{\Sigma_0}^2\Lambda_0}{4\tau(\Lambda_0 \tau - 15{\cal
K})}\,.
\ee

In such a cosmological setting, the Hubble parameter is
\be \label{42}
H = H(t) \equiv \frac{\dot{a}(t)}{a(t)} \equiv
H(\tau) = \frac{\Sigma_0}{a} \frac{d a}{d\tau} =
\frac{3\Sigma_0}{2\tau}\frac{\Lambda_0 \tau - 10 {\cal
K}}{\Lambda_0 \tau - 15 {\cal K}}\,.
\ee
Interestingly, if one studies the Hamilton equations for the pure-gravity
Lagrangian (2.2) \cite{Bona04}, and if one looks for fixed points of the
resulting system, at which $a,G$ and their conjugate momenta have vanishing
time derivative, one finds again that $\Lambda$ and $G$ are related by
$\Lambda=\Lambda_{0}G^{-2}$ as in (3.25), but in a closed universe only.

When, in addition, we are in the spatially flat universe, with
${\cal K} = 0$, Eqs. (3.23), (3.24), (\ref{3.25}) and (\ref{42})
become, respectively,
\be \label{43}
a(\tau) =
\frac{2}{3\Sigma_0}\sqrt{\frac{\Lambda_0}{5}}{\tau}^{3/2}\,,\,\,\,\,
G(\tau) =
\frac{2}{\Sigma_0}\sqrt{\frac{\Lambda_0}{15}}\tau\,,\,\,\,\,
\Lambda (\tau) = \frac{15 {\Sigma_0}^2}{4{\tau}^2}\,,\,\,\,\,
H(\tau) = \frac{3\Sigma_0}{2\tau}\,. 
\ee

\subsection{Case with $J\neq 0$ and ${\cal K}=0$}

Following the same steps as above, Eq. (3.17) gives
\be \label{44}
v = v(\tau) = \frac{2J-3}{J-2}\log{\left[ \frac{J-2}{2(2J-3)}\tau
\right]^{1/3}} = \frac{6n m}{m+1}\log{\left[ \frac{m+1}{12n m}\tau
\right]^{1/3}},
\ee
because now $\tau \equiv \Sigma_J t + 6C_1$,
and Eqs. (3.19) and (3.28) yield (with $C_2 = 0$) 
\be \label{45} 
u= u(\tau) = \frac{2(2J-3)\Lambda_0}{(3J-5){\Sigma_J}^2}\left[
\frac{J-2}{2(2J-3)}\tau \right]^{\frac{3J-5}{J-2}} = \frac{12n
 m\Lambda_0}{(6n m + m + 1){\Sigma_J}^2} \left[
\frac{m+1}{12n m}\tau \right]^{\frac{6n m + m + 1}{m+1}}\,. 
\ee
Substituting them into Eqs. (3.10) and (3.11), we get 
\ba 
a=a(\tau) & = & a_0 {\tau}^{\frac{12n^2}{6n-1}}, \\ G = G(\tau) & = &
G_0 {\tau}^{2(3n-1)}\,, 
\ea 
where we have defined 
\ba 
a_0 & \equiv
& a_0(n) \equiv \left[ \frac{12^{-\frac{6n+1}{6n-1}}}{12n-1}\left(
\frac{6n-1}{n}
\right)^{\frac{12n}{6n-1}}\frac{\Lambda_0}{{\Sigma_J}^2} \right]^n \,,\\
G_0 & \equiv & G_0(n) \equiv \left[ \frac{(6n-1)^2\Lambda_0}{12n^2
(12n-1){\Sigma_J}^2} \right]^{3n-1}\,. 
\ea
Once more, we find that all expressions in this case are
generalizations of the ones we have worked out above in the
case with $J = 0$, with the already mentioned caution as to
the terms with ${\cal K}\not =0$. As a matter of fact, when
$n=1/2$ we get $12n^2/(6n-1)=3/2$ and
$a_0=2\sqrt{\Lambda_0}/(3\sqrt{5}\Sigma_0)$ for the scale factor
$a(\tau)$, and $2(3n-1)=1$,
$G_0=2\sqrt{\Lambda_0}/(\sqrt{15}\Sigma_0)$ for the gravitational
coupling $G(\tau)$. Also, note that in the expressions for $a$ and
$G$ we have preferred to use the $n$ parameter only.

There are other non-declared constraints on the allowed values of
the $n$ parameter. In fact, so far we have simply stressed
that $n$ should be non-vanishing. But, in our calculations, we have
instead used more stringent constraints on its possible values.
Already the introduction of $m$ in fact gives $n\neq 1/3$;
furthermore, now, we have also to set $n\neq 1/12, 1/6$ so as to
obtain meaningful expressions above. As to the
influence of this on the values of the interaction parameter $\mu$
occurring in the Lagrangian (2.2), it
has also to be $\mu \neq 2/27$ (we already found that $\mu$ should
be $\neq 0,2/3$, by virtue of $J\neq 1,3/2$).

\section{Noether symmetries and RG evolution}

In this paper we have not considered $G$ and $\Lambda$ in the RG
equations as external fields, and we have therefore to check whether
our solutions can represent some RG trajectory which can always be
described as a given $\Lambda=\Lambda(G)$ law, as we said in the
introduction. Indeed, by taking into account Eqs. (3.31) and (3.33), let  
us exhibit explicitly the behaviour of the function $\Lambda = \Lambda
(\tau)$, which is peculiar of the pure-gravity regime of the
universe when $J\neq 0$ and ${\cal K}=0$: 
\be 
\label{3.34} 
\Lambda
= \Lambda (\tau) = \Lambda_0 G^{2(J-1)} = \frac{12n^2
(12n-1){\Sigma_J}^2}{(6n-1)^2}{\tau}^{-2}\,, 
\ee 
since $J-1=-m=-1/[2(3n-1)]$. This, of course, reduces to what has been
found before for $J = 0$ (and ${\cal K}=0$).

From (4.1) we have 
\begin{equation}
\Lambda G^{2(1-J)}={\rm constant},
\label{(4.2)}
\end{equation}
therefore our solutions do not describe the evolution near the fixed
point for which $\Lambda G={\rm constant}$, unless $J=1/2$. On the other
hand, it is not difficult to understand the scaling law (4.2) in terms
of the phase diagram described in \cite{Rreut04}. In fact, in the
cross-over region between the non-Gaussian and the Gaussian fixed point,
the dimensionful coupling constant scales approximately as a vacuum
energy $\Lambda \sim k^{4}$, while the dimensionful Newton constant
spends most of the RG-time in approaching the Gaussian fixed point.
With the notation of figure 4 of \cite{Rreut04}, we are approaching the
point $T$ of the phase diagram. Therefore $G \sim k^{-2}$ and 
$\Lambda G^{2} \sim {\rm constant}$. Since we know from the RG-evolution 
that the anomalous dimension in this region is approximately
vanishing, we conclude that only the $J \sim 0$ values are consistent
with the RG evolution in this region. Other values of $J$ can still 
represent viable cosmologies, but their physical meaning in terms of
the RG flow is still obscure to us. On the other hand, the cosmology 
described by the J=0 solution can be thought of as an effective description
of the Universe near the Planck era, before the transition to classical
cosmology of Fredmann--Lemaitre--Robertson--Walker type, when the seeds
for cosmological perturbations are generated.

\section{The exponent $p$ and parameter $\mu$ as functions of $J$}

In this section we begin by paying particular attention to the possibility
of obtaining inflation from our model. As is well known, the problem
of inflationary models is far from being solved. The use of a scalar
field, together with a wide range of possible potentials and couplings,
allows for good agreement with observation and theoretical requirements,
but the nature of the scalar field remains completely unknown. As we said
above, our approach, both for the choice of treating $G$ as a variable,
as well as for the Noether Symmetry Approach, 
seems to us more natural. As we show below, we
obtain a power-law inflation with arbitrarily large exponent, which 
seems very encouraging. Of course, much work should still be done. In
particular, the problem of graceful exit from inflation and that of the
perturbations' spectrum should be faced, but this 
seems a rather difficult task.

While in \cite{Bona04} a power-law behaviour for the scale factor
was guessed to solve the equations, this now has been obtained exactly
and, in a sense, more generally from them. To better
understand what we mean by this, let us first recall that, when
${\cal K}=0$, in \cite{Bona04} it was supposed $a=At^{\alpha}$
for the scale factor; with arbitrary $A$, this gives $\alpha_{\pm}
= (3\pm \sqrt{9+12{\xi}^2 \lambda_{\star}})/6$, which is closely
connected to the hypothesis of being in the neighbourhood of an ultraviolet
fixed point $(g_{\star},\lambda_{\star})$, hence constraining the
evolution of $G$ and $\Lambda$ to be given by 
\be \label{3.35} 
G=G(t) = g_{\star}{\xi}^{-2}t^2\,,\,\,\,\,\, \Lambda = \Lambda (t) =
\lambda_{\star}{\xi}^2 t^{-2}\,\,\,\, \Rightarrow \,\,\,\,
G\Lambda = g_{\star}\lambda_{\star} \equiv {\rm constant} \,. 
\ee 
This anyway also allows arbitrarily large values for ${\alpha}_{+}$,
$\xi$ being undetermined, while constraining the possible values
of the interaction parameter to be $\mu_{\pm} = 3\alpha_{\pm}/2 =
(3\pm \sqrt{9+12{\xi}^2 \lambda_{\star}})/4$ \cite{Bona04}. With
$\lambda_{\star} > 0$, we then have power-law inflation for the
``+" sign and a possible solution of the horizon problem.

Now, to discuss what we have instead found here, let us
first of all note that in our derivation we have preferred so far
to restrict our attention to the particular case with $C_2=0$,
hence simplifying the final expressions, in order to offer an easier
way to discuss the results. In the case with ${\cal K}=0$ we can
use the generic $J$ formulae, which also include that situation.
The constant $\Sigma_J$ can be set equal to $1$, for example, hence
arbitrarily fixing the time scale. Let us also set $C_1 =0$ in
order to get $a(0)=0$, which is of course more delicate, even if
using $\tau$ instead of $t$ makes it less problematic. What is 
more important is $C_2$, indeed, while we can also set
$\Lambda_0 =1$. This latter in fact appears as a factor in the
expression of $u$, and suitably redefining $C_2$ we simply have a
constant multiplying the scale factor $a$ that can be fixed
arbitrarily. All this changes much when ${\cal K} \neq 0$; in that
case the above factor depends on $J$. Thus, without need to
introduce the parameters $n$ and $m$, we can look at the
expressions where $J$ occurs, just to take into account
possible degenerations for some special values of $J$.

\begin{figure} 
\centering
\resizebox{8.5cm}{!} {\includegraphics{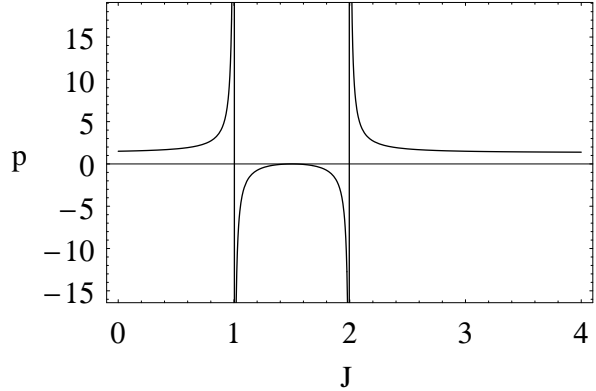}} \caption{The
exponent $p = p(J)$.} \label{exponent}
\end{figure}

\begin{figure} 
\centering \resizebox{8.5cm}{!}
{\includegraphics{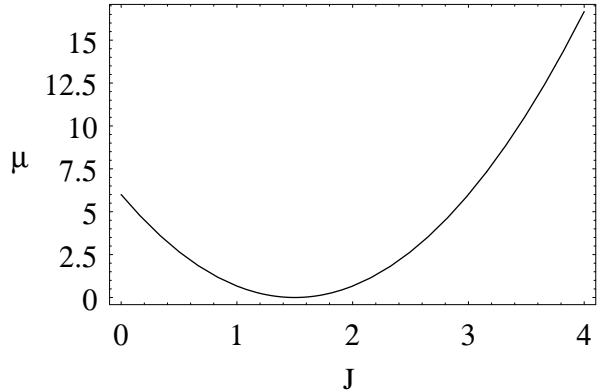}} \caption{The parameter $\mu = \mu
(J)$.} \label{parameter}
\end{figure}

With $C_2 =0$, the behaviour of the scale factor (given in
Eq. (3.31) using $n$) also describes the general asymptotic trend
of $a$ at large $\tau$. We find a power-law behaviour
$a \sim {\tau}^p$ without strictly imposing to be near a
fixed point. The exponent is 
\be \label{3.35bis} 
p \equiv
\frac{(3-2J)^2}{3(J^2 -3J +2)}\,, 
\ee 
whose plot is shown in figure \ref{exponent}.
We find inflation only when $J<1$ or $J>2$, $p$ being always
$>1$ in these ranges; in particular, when $J$ is near the values
$1$ and $2$, the exponent can assume arbitrarily large values,
hence giving a very strong power-law behaviour. On the other
hand, the interaction parameter $\mu$ is plotted in figure
\ref{parameter} as a function of $J$, hence showing that both $p$ and
$\mu$ assume symmetric values for $J<1$ and $J>2$.
By virtue of the physical character of the $\mu$ parameter, we
might think to limit our considerations to the case with $J>2$, hence
avoiding to take $J<0$ into account. But, as we will see later,
the behaviour of $G$ is strongly affected by such a choice. In any
case, there seem to exist two disconnected families of solutions,
which can be probably seen as resulting from the ${\dot{G}}^2$ term in the
equations, producing roots with different signs. As to the
function $G = G(\tau)$, on setting $C_2 = 0$  we still have a
power-law behaviour, with exponent 
\be \label{3.35bis1}
p' \equiv \frac{1}{1-J}\,. 
\ee 
This leads to two alternative cases, one in
which $G$ increases and another in which $G$ decreases. In any
case, for $J<1$ and $C_2 \neq 0$, $G$ diverges as $\tau
\rightarrow 0$. A special situation is obtained when $p^{\prime}
=2$, that is $J=1/2$; this is an inflationary case, with $p=16/9$,
and we discuss it in the following section. In the $J=0$ case, the
exponent $p$ is instead fixed to be $3/2$ (as can be seen in Eqs.
(3.23) and (\ref{43}), hence giving only a soft acceleration for the
early universe.

\section{Concluding remarks and open problems}

Even though the proximity of the fixed point is somehow partially
guessed at the beginning (for physically justifying our Lagrangian
formulation), we have, anyway, never used expressions like those
in Eq. (\ref{3.35}), but we have gained information on the
$\Lambda (G)$ form by the existence of the assumed Noether
symmetry. Here, we see that the choice we have consequently done
on that concretely limits the generality of our solution. Thus, we
discover that, when $J=0$ and ${\cal K}=0$, one has 
\be
\label{3.36} 
G(\tau)\Lambda (\tau) =
\frac{\sqrt{15\Lambda_0}\Sigma_0}{2}{\tau}^{-1}\,, 
\ee 
while, for $J\neq 0$ and ${\cal K}=0$, we get 
\be 
\label{3.37} 
G(\tau)\Lambda
(\tau) = \Lambda_0 {G_0}^{\frac{3n-2}{3n-1}}{\tau}^{2(3n-2)}\,.
\ee 
In both cases, therefore, we do not find the typical constant
behaviour for the product $G\Lambda$ in the neighbourhood of the
fixed point. As a matter of fact, in Ref. \cite{Bona04} the
behaviour we show in Eq. (4.1) is imposed from the beginning in
the equations from the RG-improved Lagrangian for pure gravity,
while we indeed solve them exactly by imposing a particular choice
for the product $G\Lambda$ inspired by the existence of the
Noether symmetry for the Lagrangian (see comments following Eq. 
(3.26)). It is interesting, anyway,
that, in this respect, what we find in this paper can indeed be
reduced to the results in Ref. \cite{Bona04} only when $J=1/2$ (a
single possible value, from which $n=2/3$ and $\mu = 8/3$), hence
implying $G\Lambda = \Lambda_0 \equiv g_{\star}\lambda_{\star}$.
This also fixes the power-law behaviour of the scale factor to be
$a(\tau) \sim {\tau}^{16/9}$, as said above. The fact that
$\mu$ has now to be equal to $8/3$, while in Eq. (\ref{3.35}) it
can assume any value (the $\alpha$ parameter being, indeed, free
to take any value), means that, even though the Noether symmetry
might also be seen as a sort of peculiarity of the physical
situation we have described in this paper, there is no
easy extension of the solutions in Ref. \cite{Bona04} to ours,
without a very sharp fine tuning.

As a further remark, let us anyway note that, for the scale factor
behaviour, in Ref. \cite{Bona04} the case with ${\cal K} = \pm 1$
yields $\alpha = 1$, i.e., $a=At$, while we have here found
(considering the situation with $J=0$ and generic ${\cal K}$) the
expression in Eq. (3.23). These scale factors are not in
contradiction, indeed, even if the one we have here calculated
contains explicitly the ${\cal K}$ contribution. In fact,
by virtue of Eqs. (3.24) and (\ref{3.25}) we get
\be \label{3.38}
G(\tau)\Lambda (\tau) = \frac{\sqrt{15}\Sigma_0
\Lambda_0}{2\sqrt{\tau (\Lambda_0 \tau - 15{\cal K})}}\,.
\ee

All these considerations seem to support the idea that the 
model discussed in this paper should indeed concern an era just
subsequent to the one characterized by the ultraviolet fixed point we
talked about at the beginning, an era in which matter fields (and
other non-gravitational fields) can still be neglected
with respect to the gravity contribution. At that time,
then, the time variation of $G$ and $\Lambda$ were as
relevant for the cosmic evolution as they were in the earlier
stage, whose description has been given elsewhere (see Ref.
\cite{Bona04} and references therein). One of the important 
open issues concerns, therefore, the way a link might be
found between such two different physical situations, which
demands much more future work. 

On the other hand, our new results about the functional relation between
$\Lambda$ and $G$, as in (4.2), jointly with the power-law inflationary
solutions that we have described so far, look very promising on
the path towards further insight. Encouraging evidence is therefore
emerging in favour of models with variable $G$ and $\Lambda$ being able
to lead to a deeper understanding of modern cosmology
\cite{Barv93, Baue05, Laus05, Bona03, Scud06}.

\acknowledgments
The authors are grateful to the INFN for
financial support. The work of G. Esposito has been partially
supported by PRIN {\it SINTESI}.


\begin{references}
\bibitem{Berg02}
Berges J, Tetradis N and Wetterich C 2002 {\it Phys. Rep.} {\bf
363} 223
\bibitem{Wett01}
Wetterich C 2001 {\it Int. J. Mod. Phys.} A {\bf 16} 1951
\bibitem{Reut94}
Reuter M and Wetterich C 1994 {\it Nucl. Phys.} B {\bf 427} 291
\bibitem{Reut98}
Reuter M 1998 {\it Phys. Rev.} D {\bf 57} 971
\bibitem{Laus02a}
Lauscher O and Reuter M 2002 {\it Phys. Rev.} D {\bf 65} 025013
\bibitem{Soum99}
Souma W 1999 {\it Prog. Theor. Phys.} {\bf 102} 181
\bibitem{Laus02b}
Lauscher O and Reuter M 2002 {\it Class. Quantum Grav.} {\bf 19}
483
\bibitem{Reut02b}
Reuter M and Saueressig F 2002 {\it Phys. Rev.} D {\bf 65} 065016
\bibitem{Bona05}
Bonanno A and Reuter M 2005 {\it J. High Energy Phys.}
JHEP02(2005)035
\bibitem{Nied03}
Niedermaier M 2003 {\it Nucl. Phys.} B {\bf 673} 131
\bibitem{Nied02}
Niedermaier M 2002 {\it J. High Energy Phys.}
JHEP12(2002)066
\bibitem{Forg02}
Forgacs P and Niedermayer M 2002 hep-th/0207028
\bibitem{Bona02b}
Bonanno A and Reuter M 2002 {\it Phys. Rev.} D {\bf 65} 043508
\bibitem{Bona99}
Bonanno A and Reuter M 1999 {\it Phys. Rev.} D {\bf 60} 084011
\bibitem{Bona00}
Bonanno A and Reuter M 2000 {\it Phys. Rev.} D {\bf 62} 043008
\bibitem{Bona02a}
Bonanno A and Reuter M 2002 {\it Phys. Lett.} B {\bf 527} 9
\bibitem{Reut04}
Reuter M and Weyer H 2004 {\it Phys. Rev.} D {\bf 69} 104022;
{\it Phys. Rev.} D {\bf 70} 124028
\bibitem{Bona04}
Bonanno A, Esposito G and Rubano C 2004 {\it Class. Quantum Grav.}
{\bf 21} 5005
\bibitem{Tsam93}
Tsamis N C and Woodard R P 1993 {\it Phys. Lett.} B {\bf 301} 351
\bibitem{deritis90}
de Ritis R, Marmo G, Platania G, Rubano C, Scudellaro P, and
Stornaiolo C 1990 {\it Phys. Rev.} {\bf D 42} 1091
\bibitem{cap96}
Capozziello S, de Ritis R, Rubano C, and Scudellaro P 1996 {\it
Rivista del Nuovo Cimento} {\bf 19(4)} 1
\bibitem{Rreut04}
Reuter M and Weyer H 2004 {\it J.Cosmol. Astropart. Phys.} 
JCAP12(2004)001
\bibitem{deri90}
de Ritis R, Marmo G, Platania G, Rubano C, Scudellaro P and
Stornaiolo C 1990 {\it Phys. Lett.} A {\bf 149} 79
\bibitem{deri95}
de Ritis R, Rubano C and Scudellaro P 1995 
{\it Europhys. Lett.} {\bf 32} 185
\bibitem{Barv93}
Barvinsky A O, Kamenshchik A Yu and Karmazin I P 1993 
{\it Phys. Rev.} D {\bf 48} 3677
\bibitem{Baue05}
Bauer F 2005 gr-qc/0512007
\bibitem{Laus05}
Lauscher O and Reuter M 2005 hep-th/0511260
\bibitem{Bona03}
Bonanno A, Esposito G and Rubano C 2003 {\it Gen. Rel. Grav.} {\bf 35}
1899
\bibitem{Scud06}
Bonanno A, Esposito G, Rubano C and Scudellaro P 2006 astro-ph/0612091
\end{references}
\end{document}